\documentclass[12pt,preprint]{aastex}
\usepackage{graphicx}
\begin{document}
\title{Are GRB shocks mediated by the Weibel instability?}

\author{Yuri Lyubarsky, David Eichler}
\affil{Physics Department, Ben-Gurion University, P.O.B. 653,
Beer-Sheva 84105, Israel; lyub,eichler@bgumail.bgu.ac.il}

\begin{abstract}
It is estimated that the Weibel instability is not generally an
effective mechanism for generating ultrarelativistic astrophysical
shocks. Even if the upstream magnetic field is as low as in the
interstellar medium, the shock is mediated not by the Weibel
instability but by the Larmor rotation of protons in the
background magnetic field. Future simulations should be able to
verify or falsify our conclusion.
\end{abstract}


\maketitle

\section{Introduction}
There is large literature on gamma-ray burst (GRB) afterglows
based on the assumption that the X-ray, optical and radio
afterglows are the synchrotron emission from relativistic
electrons Fermi accelerated at the forward shock of the blast wave
(see recent reviews by M\'{e}sz\'{a}ros 2002; Zhang \&
M\'{e}sz\'{a}ros 2004). A longstanding difficulty with this
assumption has been the inferred magnetic field needed to fit the
afterglow data typically requires that the magnetic energy density
exceeds by many orders of magnitude that which would be expected
from the shock compression of the interstellar magnetic field of
the host galaxy (Gruzinov \& Waxman 1999; Gruzinov 2001). Many
authors have therefore assumed that the shock somehow manufactures
field energy to meet this requirement, but no convincing mechanism
has been proposed to date. One mechanism discussed is the Weibel
instability (Medvedev \& Loeb 1999, Silva et al. 2003; Frederiksen
et al. 2004; Jaroschek et al. 2005; Medvedev et al. 2005; Kato
2005; Nishikawa et al. 2005), which has the fastest growth rate
and produces relatively strong small scale magnetic field even in
an initially non-magnetized plasma. It is expected that the
thermalization of the upstream flow could occur via scattering of
particles on the magnetic fluctuations. In the electron-positron
plasma, the instability does generate the magnetic field at about
10\% of the equipartition level and does provide the shock
transition at the scale of a dozen of electron inertial length
(Spitkovsky 2005). However, simulations of colliding
electron-proton flows show that while the electrons are readily
isotropized, the protons acquire only small scattering in angles
after passing the simulation box (Frederiksen et al. 2004).
How long the field persists after the shock is also an important
question (Gruzinov 2001) but here we discuss whether the Weibel
instability can even cause the shock in the first place.

Moiseev \& Sagdeev (1963, see also Sagdeev 1966) analyzed the
structure of the nonrelativistic Weibel driven shock and found
that the width of the shock transition should be very large
because electrons easily screen proton currents thus suppressing
development of the instability. Failure of the Weibel instability
to preempt other shock mechanisms, except for very large Alfven
Mach numbers, has been discussed in the context of nonrelativistic
shocks by Blandford and Eichler (1987).  Here we estimate the
width of the Weibel driven shock in the ultra-relativistic
electron-proton plasma. Although based on a number of physical
assumptions  about the behavior of the plasma parameters at the
non-linear stage of the Weibel instability, such analytical
scalings are necessary in any case, because, by evident reasons,
simulations of plasmas are possible only with artificially low
proton-to-electron mass ratios (e.g., Frederiksen et al. (2004)
took $m_p/m_e=16$). In this paper, we present the parameters in
physically motivated dimensionless form and we believe that our
assumptions could be checked by numerical simulations. Only by
combining numerical simulations with analytical scalings can we
achieve reliable conclusions about the properties of real shocks.

As a model for the shock formation, we consider collision of two
oppositely directed plasma flows. Eventually two diverging shocks
should be formed with plasma at rest between them. However at the
initial stage, the two flows interpenetrate each other exciting
turbulent electro-magnetic fields. Particles are eventually
thermalized by scattering off these turbulent fields. As electrons
are thermalized relatively rapidly, we consider development of the
Weibel instability in two proton beams propagating through
relativistically hot isotropic electron gas. We estimate the
proton isotropization length in such a system and conclude that
the Weibel-mediated shocks are so wide that even in the
interstellar medium, the shock should be formed at the scale of
the Larmor radius of the proton in the background magnetic field.

The article is organized as follows. In sect.2, we find the growth
rate of the proton Weibel instability. Saturation of the
instability is considered in sect.3. In sect.4, we exploit the
obtained results in order to estimate the width of the
Weibel-mediated shock transition. Section 5 contains the
discussion.

\section{The Weibel instability}
Let us consider two proton beams of equal strength propagating in
opposite directions along the $z$-axis. For the sake of
simplicity, let us adopt the waterbag distribution function:
 \begin{equation}
F_p(\mathbf{p})=\frac 1{2\pi p_{\perp 0}^2}
[\delta(p_{z}-p_{\parallel 0})+\delta(p_{z}+p_{\parallel
0})]\Theta(p_{\perp 0}^2-p_{\perp}^2);\label{distr}
\end{equation}
where $\Theta(x)$ is the Heaviside step function,
$p_{\perp}=\sqrt{p_x^2+p_y^2}$. Assume that the beams propagate
through an isotropic relativistically hot electron plasma with the
distribution function $F_e(p)$.

This configuration is known to be unstable because a small
transverse magnetic field
$\mathbf{B}=B\mathbf{\widehat{x}}\exp{(-\imath\omega+\imath ky)}$
would drive the oppositely moving protons into current layers of
opposite sign, which reinforce the initial field (see, e.g., Fig.1
in Medvedev \& Loeb 1999). Evolution of the electromagnetic field
is governed by Maxwell's equations
\begin{equation}
\mathbf{\nabla\times E}=-\frac{\partial\mathbf{B}}{\partial
t};\quad \mathbf{\nabla\times
B}=\frac{\partial\mathbf{E}}{\partial t}+4\pi\mathbf{j};
\end{equation}
which are written in Fourier space as
\begin{equation}
kE=\omega B;\quad \imath(\omega E-kB)=4\pi j.\label{Maxwell}
\end{equation}
Note that only $z$ components of $\mathbf{E}$ and $\mathbf{j}$ are
present in this configuration. The current density, $j$, may be
found from a solution to the linearized Vlasov equation
\begin{equation}
\imath(\omega-kv_y)\delta F_{p,e}=\pm
e\left[(E-v_yB)\frac{\partial F_{p,e}}{\partial
p_z}+v_zB\frac{\partial F_{p,e}}{\partial p_y}\right]
\end{equation}
as
\begin{equation}
j=en\int v_z(\delta F_{p}-\delta F_e)d\mathbf{p}.\label{current}
\end{equation}
As usual (e.g., Krall \& Trivelpiece 1973), the condition for the
set of equations (\ref{Maxwell}--\ref{current}) to have a nonzero
solution yields the dispersion equation
\begin{equation}
k^2=\omega^2[1+\chi_e(\omega,k)+\chi_p(\omega,k)];\label{disp}
\end{equation}
where
\begin{equation}
\chi_{\alpha}=\frac{4\pi e^2n}{\omega^2}\int v_z
 \left\{\frac{\partial F_{\alpha}}{\partial p_z}+
 \frac{kv_z}{\omega+\imath 0-kv_y}\frac{\partial F_{\alpha}}{\partial p_y}\right\}
 d\mathbf{p}\label{suscept}
\end{equation}
is the susceptibility of the plasma species $\alpha$.

Substituting the proton distribution function (\ref{distr}) into
Eq.(\ref{suscept}) yields
$$
\chi_p(\omega,k)=-\frac{2\Omega_{pp}^2}{\omega^2}\left\{
 \frac 1{m_p\gamma+\sqrt{m_p^2\gamma^2-p^2_{\perp0}}}\left(m_p\gamma-\frac{p_{\parallel0}^2}{\sqrt{m_p^2\gamma^2-p^2_{\perp0}}}\right)
 \right.$$\begin{equation}\left.
 +\frac{p_{\parallel0}^2}{p^2_{\perp0}}
 \left(\frac{\omega
 m_p\gamma}{\sqrt{\omega^2m_p^2\gamma^2-k^2p_{\perp0}^2}}-1\right)\right\};
\label{psuspect}\end{equation}
 where $\Omega_{pp}\equiv\sqrt{4\pi e^2/m_p\gamma}$ is the
 relativistic proton plasma frequency,
 $\gamma=\sqrt{1+(p_{\perp0}^2+p_{\parallel0}^2)/m_p^2}$.
 Below only the strongly anisotropic,  highly relativistic
 case is considered, $p_{\perp}\ll p_{\parallel}$, $\gamma\gg 1$. Then one can neglect
 the first term in the curly brackets.

 The susceptibility of isotropic electrons is written as
$$
\chi_e(\omega,k)=\frac{4\pi e^2n}{\omega}
 \int\frac{v_z}{\omega+\imath0 -kv_y}\frac{\partial F_e}{\partial
 p_z}d\mathbf{p}
 $$$$=\frac{4\pi e^2n}{\omega}
 \int\frac{v\sin^2\theta\cos^2\varphi}{\omega+\imath0 -kv\cos\theta}
 \frac{dF_e}{dp}p^2dpd\cos\theta d\varphi
 $$\begin{equation}
=\frac{4\pi^2 e^2n}{\omega}\left\{
\wp\int\frac{v(1-x^2)}{\omega-kvx}\frac{dF_e}{dp}p^2dxdp-
\frac{\pi\imath}k\int
\left(1-\frac{\omega^2}{k^2v^2}\right)p^2\frac{dF_e}{dp}dp\right\}.
\label{esusp}\end{equation} Here we used the spherical coordinates
in the momentum space and the Plemeli formula. The Weibel
instability operates in the low-frequency limit, $\omega\ll k$. In
this case the imaginary part of $\chi_e$ dominates because at
$\omega\to 0$, the principal value of the integral in $x$ goes to
zero. The physical reason is that only electrons with small $v_y$
contribute to the susceptibility because other electrons "see"
rapidly oscillating field as they move in $y$ direction  over a
distance larger than $1/k$  for the time $\sim 1/\omega$.  Now one
can write
\begin{equation}
\chi_e(\omega,k)=\imath\frac{\pi\Omega_{pe}^2}{4k\omega};
\end{equation}
where $\Omega_{pe}^2=4\pi e^2n/T$, $1/T\equiv 8\pi\int F_epdp$.
The parameter $T$ is equal to the electron temperature if the
electron spectrum is Maxwellian and $T\gg m_e$. Note that in their
analysis of the proton Webel instability, Wiersma \& Achterberg
(2004) erroneously used the expression
$\chi_e=-\Omega_{pe}^2/\omega^2$, which is valid only in the high
frequency limit, $\omega\gg k$, and therefore is irrelevant to the
case of interest.

Now one can write the dispersion equation (\ref{disp}) in the
low-frequency limit:
\begin{equation}
k^2+\frac{2\Omega_{pp}^2p_{\parallel0}^2}{p^2_{\perp0}}
\left(\frac{\omega
m_p\gamma}{\sqrt{\omega^2m_p^2\gamma^2-k^2p_{\perp0}^2}}-1\right)
-\imath\frac{\pi\Omega_{pe}^2\omega}{4k}=0.\label{disp1}
\end{equation}
 In the limit of negligible angular spread
of the proton beams, $p_{\perp0}\to 0$, the dispersion equation is
reduced to a simple cubic equation
\begin{equation}
\frac{\alpha}{\kappa^3}x^3+x^2-1=0;\label{disp2}
\end{equation}
where $\alpha\equiv \pi\Omega_{pp}^2/4\Omega_{pe}^2=\pi
m_p\gamma/4T$, $\kappa\equiv k/\Omega_{pp}$, $x\equiv
-\imath\omega/\Omega_{pp}$. The system is unstable provided $\Re
x>0$. In the small- and long-wavelength limits, the growth rate
$\eta\equiv\Im\omega=x\Omega_{pp}$ is found as
\begin{equation}
\eta=\left\{\begin{array}{ll}k\alpha^{-1/3}; &
k\ll\alpha^{1/3}\Omega_{pp};
\\ \Omega_{pp}; & k\gg\alpha^{1/3}\Omega_{pp}.
\end{array}\right.\label{growthrate}
\end{equation}
The full solution to Eq.(\ref{disp2}) is shown in Fig.1. One can
see that if $T\sim\gamma m_e$, as one can expect within the shock
structure, the most unstable are short-wave perturbations,
$k\gtrsim (m_p/m_e)^{1/3}\Omega_{pp}$.

\begin{figure*}
\includegraphics[width=10 cm,scale=0.8]{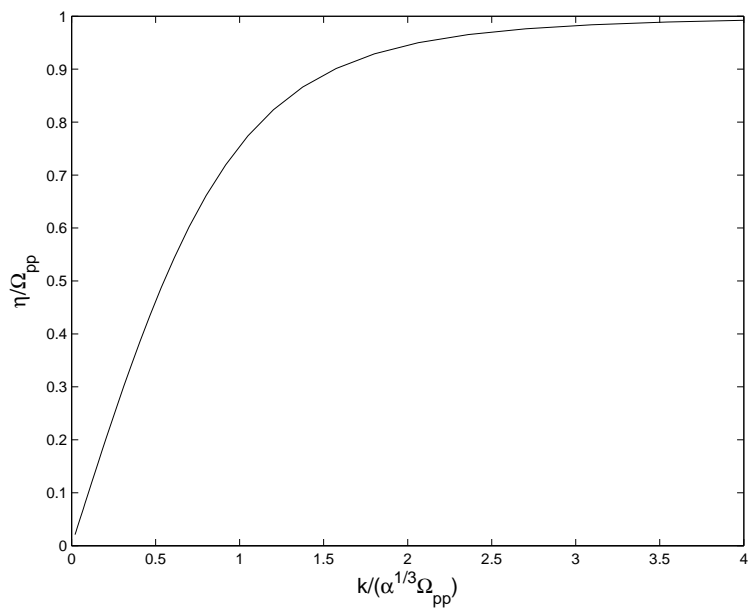}
\caption{Growth rate of the proton Weibel instability; cold beams.}
\end{figure*}
\begin{figure*}
\includegraphics[width=10 cm,scale=0.8]{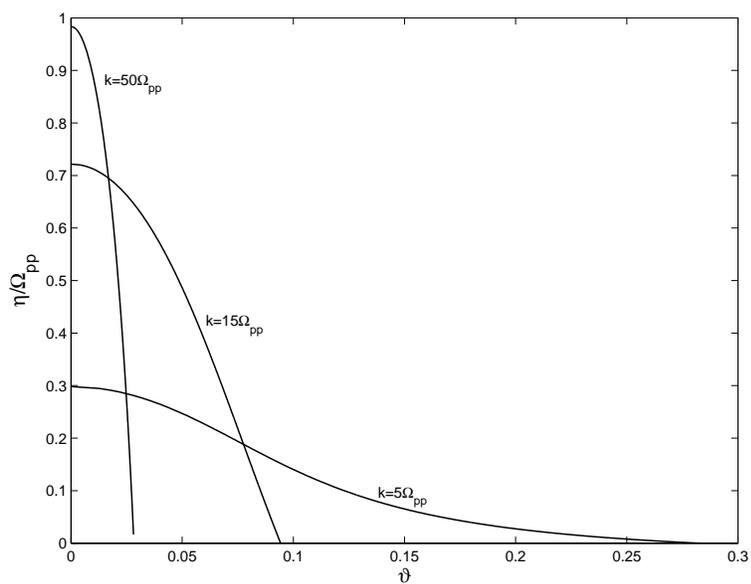}
\caption{Dependence of the growth rate on the angular spread of
the proton beams $\vartheta\equiv p_{\perp0}/p_{\parallel0}$ at
$T=m_e\gamma/3$}
\end{figure*}

When the angular spread of the beams increases, the growth rate of
the instability decreases (Fig.2). The threshold of the
instability may be easily found by substituting $\omega=0$ into
Eq.(\ref{disp1}); this yields
\begin{equation}
\frac{p_{\perp0}}{p_{\parallel0}}=\frac
{\sqrt{2}\Omega_{pp}}k.\label{threshold}
\end{equation}
So at small wavelengths, where the growth rate is maximal, the
instability stops after a small spread in the angular velocity
distribution is achieved.

\section{Stabilization of the Weibel instability}
The dispersion relation (\ref{disp}) was found assuming that the
particles are nonmagnetized, i.e. their trajectories are nearly
straight. The instability saturates when this condition is
violated either for protons or for electrons. As a result of the
instability, current filaments are formed along the direction of
the proton motion. The magnetic field forms a sort of cocoon
around these filaments and eventually traps protons within the
filaments; then the instability stops (Yang, Arons \& Langdon
1994; Wiersma \& Achterberg 2004). The quiver motion of a proton
within the current filament may be described by the linearized
equation
\begin{equation}
m_p\gamma\frac{d^2\xi}{dt^2}=ev_zB(\xi ,t);\label{quiver}
\end{equation}
where $\xi$ is the proton displacement in the transverse
direction. Near the axis of the filament, the magnetic field may
be written as $B\approx k\xi B_0$, where $B_0\propto\exp{\eta t}$
is the amplitude of the perturbation. Then Eq.(\ref{quiver})
describes oscillations in the transverse direction with the
frequency $\omega_0=\sqrt{eB_0k/m_p\gamma}$ and the (growing)
amplitude $\xi=eB_0/(\gamma m_p\eta^2)$. The proton is trapped
within the filament when the oscillation amplitude becomes less
than $1/k$ or, which is the same, frequency $\omega_0$ exceeds the
growth rate of the instability, $\eta$. This occurs when the
magnetic field reaches the value
\begin{equation}
B_{\rm trap}=\frac{m_p\gamma\eta^2}{ek}=
\frac{m_p\gamma}e\left\{\begin{array}{ll}k\alpha^{-2/3}; &
k\ll\alpha^{1/3}\Omega_{pp};
\\ \Omega_{pp}^2/k; & k\gg\alpha^{1/3}\Omega_{pp}.
\end{array}\right.\label{trapB}
\end{equation}

One can consider electrons as non-magnetized if their Larmor
radius exceeds the characteristic scale of the unstable
perturbation (Moiseev \& Sagdeev 1962; Sagdeev 1966). This
condition is violated when the field reaches the value
\begin{equation}
B_{\rm fr}=\frac{Tk}e.\label{frozB}
\end{equation}
Then the magnetic field becomes frozen into the electrons and the
magnetic flux does not grow more.

\begin{figure*}
\includegraphics[scale=0.8]{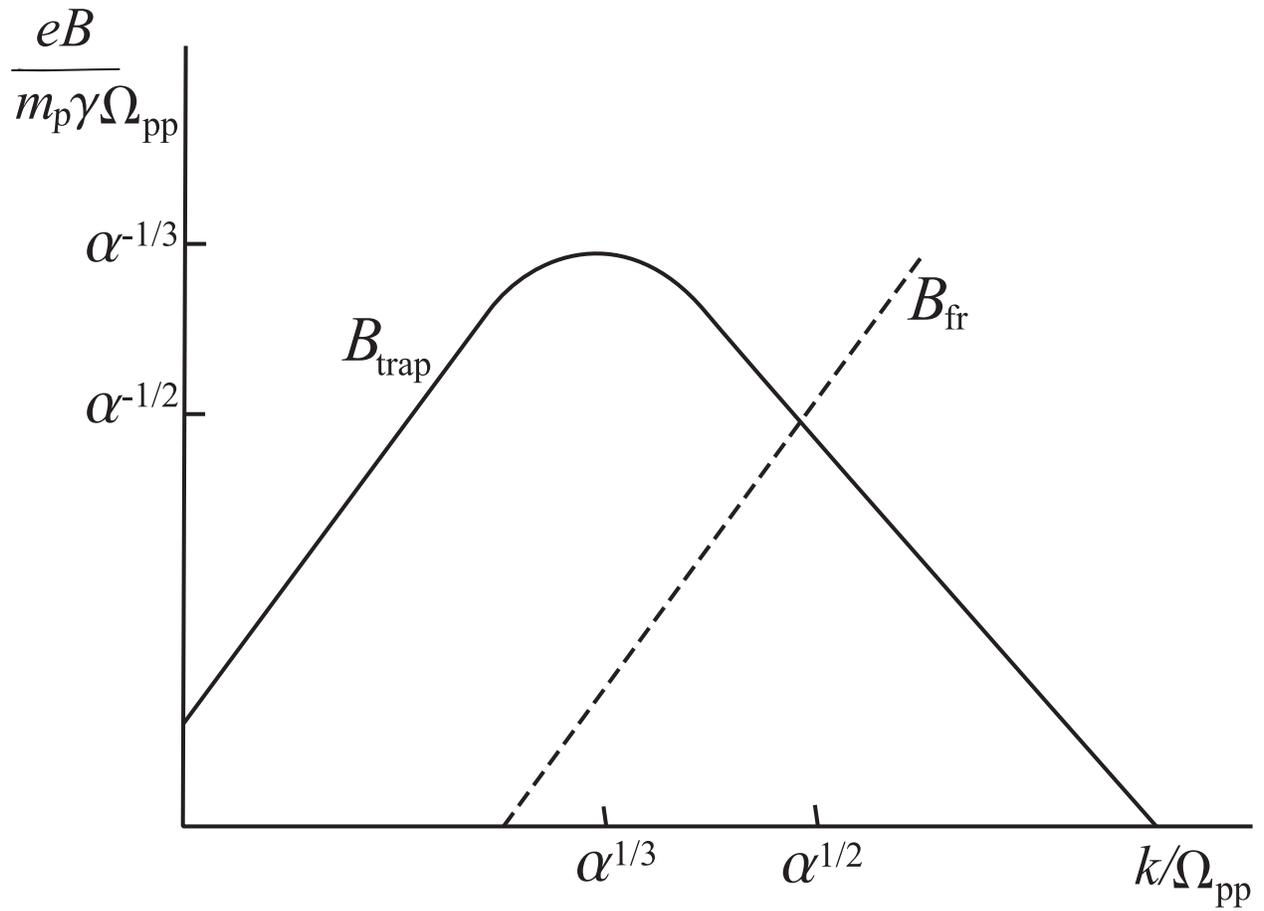}
\caption{Saturation magnetic field as a function of the wave
number. The limit $B_{\rm trap}$ due to the proton trapping is
shown by solid line; the limit $B_{\rm fr}$ when the field becomes
frozen into electrons is shown by dashed line}.
\end{figure*}

The obtained limits on the magnetic field are shown in Fig.3. The
instability develops until the magnetic field $B_{sat}={\rm
min}(B_{\rm fr},B_{\rm trap})$ is reached. One can see from Fig.3
that the maximal field is achieved in perturbations with the wave
number
\begin{equation}
k_0=\alpha^{1/2}\Omega_{pp}=\sqrt{\frac{\pi}4}\Omega_{pe};\label{k0}
\end{equation}
the corresponding wavelength is of the order of the inertial
length of electrons. The energy of the generated field scales as
the energy of electrons:
\begin{equation}
\frac{B^2}{8\pi}=\frac
1{8\pi\alpha}\left(\frac{m_p\gamma\Omega_{pp}}e\right)^2=\frac
2{\pi}nT.\label{B0}
\end{equation}
Within the shock structure, one can conveniently normalize the
electron temperature as
\begin{equation}
T=\tau m_e\gamma/3;\label{temp}
\end{equation}
where $\tau$ is a dimensionless parameter; $\tau=1$ means that the
average electron energy remains the same as upstream of the shock.
Now one can estimate the fraction of the upstream kinetic energy
transformed into the energy of the magnetic field as
\begin{equation}
\epsilon_B=\frac{2\tau m_e}{3\pi m_p}.\label{maznetization}
\end{equation}

\section{The width of the shock wave}

When two oppositely directed plasma streams collide, the Weibel
instability generates small scale magnetic fields; particle
scattering off these magnetic fluctuations provides an
isotropization mechanism necessary for the shock transition to
form. The electron streaming is halted easily whereas protons
still plow on through an isotropic electron gas. The shock is
formed on the scale defined by slow diffusion of protons in the
momentum space (Moiseev \& Sagdeev 1962; Sagdeev 1966).

Let us denote the transverse scale of the magnetic inhomogeneities
by $d$ and the amplitude of magnetic fluctuations by $B$. The
proton is scattered over a characteristic correlation length by
the angle
\begin{equation}
\delta\theta=\frac{eBd}{\sin\theta m_p\gamma};
\end{equation}
where $\theta$ is the angle the proton makes with the flow
direction. Here we take into account that the Weibel instability
generates strongly elongated current filaments so that the proton
passes the distance $l=d/\sin\theta$ within the same filament (of
course $l$ and $\delta\theta$ remain finite at $\theta\to 0$
however we are interested in isotropization scale, which is
determined by $\theta\sim 1$). The angular diffusion coefficient
is estimated as
\begin{equation}
D=\frac{(\delta\theta)^2}l=\frac{e^2B^2d}{\sin\theta
m_p^2\gamma^2};
\end{equation}
which yields the isotropization scale
\begin{equation}
L=\frac 1d\left(\frac{m_p\gamma}{eB}\right)^2.\label{scale}
\end{equation}
Motivated by the estimates expressed in Eqs. (\ref{k0}),
(\ref{B0}), and (\ref{temp}), we normalize the characteristic
inhomogeneity scale, $d$, and turbulent
 magnetic field amplitude, $B$,  as
\begin{equation}
B=4\xi\sqrt{\tau m_e\gamma/3}; \qquad
d=\zeta/(\alpha^{1/2}\Omega_{pp});
\end{equation}
where $\tau$, $\xi$, and $\zeta$ are dimensionless parameters. Now
the shock width may be expressed as
\begin{equation}
L=\left(\frac{3\pi m_p}{4\tau m_e}\right)^{3/2}\frac
1{\xi^2\zeta\Omega_{pp}}.\label{shockwidth}
\end{equation}
The estimated shock width, $L$, is thus very large compared to the
proton inertial length $1/\Omega_{pp}$ assuming that the electron
temperature as well as the scale and amplitude of the generated
magnetic field do not exceed significantly their fiducial values,
i.e. $\tau\sim\xi\sim\zeta\sim 1$. We now explain why this is
expected to be the case.

If
$\tau\sim m_p/m_e\gg 1$, then, although the shock width could then
be brought down to the proton skin depth, this would beg the
question of how the electrons are heated, which is merely passing
along the question of a shock mechanism. Similarly,
we expect $\xi\lesssim 1$ because we know of no reason to expect
that magnetic field would grow above the limit (\ref{B0}). On the
contrary, such a small-scale magnetic field should decay (Gruzinov
2001). There are evidences for hierarchical merging of current
filaments (Silva et al. 2003; Frederiksen et al. 2004; Medvedev et
al 2005; Kato 2005) so that $\zeta$ might exceed unity. However,
the merging should be accompanied by the field dissipation because
the magnetic field between two parallel adjacent currents change
sign and therefore should dissipate when the currents merge. So
possible increase of $\zeta$ would be compensated by decrease of
$\xi$. In any case, at the scale larger than the electron Larmor
radius, which is of the order of $\sim
1/(\alpha^{1/2}\Omega_{pp})$, the field is already frozen into the
electron gas therefore $\zeta$ could hardly ever grow
significantly beyond unity. One should also take into account that
the current filaments are unstable to a kink-like mode
(Milosavljevi\'c \& Nakar 2005), which also stimulates the field
decay. Therefore we believe that there is no reason for the
Weibel-driven shock transition to be significantly narrower than
Eq.(\ref{shockwidth}) predicts. There is, however, reason to
suspect that the transition is even more gradual than predicted by
Eq.(\ref{shockwidth}); this is decay of the small scale magnetic
field.

The highest resolution published simulations of shell collisions
with $m_p/m_e=16$ (Frederiksen et al. 2004) do show that while
electrons are readily isotropized, the proton beams achieve only a
small angular spread when passing the simulation box of the length
$40/\Omega_{pp}$. These simulations also show that the average
electron energy does not grow so that our conjecture $\tau\sim 1$
seems to be correct. The experiment duration, $120/\Omega_{pp}$,
was three times larger than the particle crossing time; by the end
of the simulations the spatial wavelength of the magnetic
fluctuations achieved one half of the width of the simulation box.
In physical units, this is written as $\lambda=20/\Omega_{pe}$;
$k=0.31\Omega_{pe}$ so that $\zeta=3$. Although one can not
exclude that the pattern growth was frustrated by the finite size
of the simulation box, we believe, by the reasons outlined above,
that $\zeta$ would not grow significantly in any case.

According to the estimate (\ref{shockwidth}), the full shock
transition is too wide to be simulated numerically even with a
moderate mass ratio $m_p/m_e> 10$. On the other hand,
 the scaling (\ref{shockwidth}) may hardly be ever applied to the
case $m_p/m_e\lesssim 10$ because it was obtained under the
assumption that the proton and electron scales are well separated,
i.e. that $\sqrt{m_p/m_e}\gg 1$. Therefore direct numerical check
of this scaling is very difficult. Nevertheless it would be very
useful to follow behavior of the parameters $\tau$, $\xi$ and
$\zeta$ in numerical simulations even with a low mass ratio. Even
2.5D simulations of the proton Weibel instability in the isotropic
electron gas would clarify the behavior of these parameters in
highly nonlinear regime.

\section{Discussion}
The estimate (\ref{shockwidth}) was obtained under the assumption
that there is no magnetic field in the upstream flow. If the flow
is magnetized, a shock transition may be formed at the scale of
the proton Larmor radius; therefore the above estimates are valid
only if $eB_0L/m_p\gamma<1$, where $B_0$ is the magnetic field in
the upstream flow. One can conveniently characterize the
magnetization of the flow by the parameter $\sigma =B_0^2/(4\pi
m_p\gamma n)$. Making use of Eq.(\ref{shockwidth}), one finds that
the shock may be driven by the Weibel instability if
\begin{equation}
\sigma<\xi^4\zeta^2\left(\frac{4\tau m_e}{3\pi
m_p}\right)^3=1.5\times
10^{-11}\xi^4\tau^3\zeta^2.\label{magnetization}
\end{equation}
 If the shock propagates through the
interstellar medium, the magnetization exceeds the right-hand side
of Eq.(\ref{magnetization}) by factor about 30: $\sigma=5\cdot
10^{-10}B^2_{-5.5}n^{-1}$, where $B=10^{-5.5}B_{-5.5}$ G is the
interstellar magnetic field, $n$ cm$^{-3}$ the gas number density.
This suggests that forward shocks that presumably produce GRB
afterglows are mediated not by the Weibel instability but by the
Larmor rotation of protons in the background magnetic field. This
does not mean that the Weibel instability does not work at all. On
the contrary, it does develop and may well create some small scale
magnetic field  that is much stronger than the background field.
The scattering off these magnetic fluctuations results in
diffusion of the protons in angles. However, because the
scattering does not manage  to isotropize the protons at the scale
less than the Larmor radius it is hard to see how the Weibel
instability could convert the kinetic energy to another forms. The
strong dependence of the estimate (\ref{maznetization}) on the
parameters $\xi$, $\tau$, and $\zeta$ makes accurate determination
of their values, presumably by simulations, crucial to solidify
this conclusion.

The estimate (\ref{magnetization}) shows that a fraction
$\epsilon_B\sim 10^{-4}$ of the total energy is converted into the
magnetic energy unless the electrons are heated additionally
within the shock structure. The generated small-scale field should
decay (Gruzinov 2001; Milosavlevi\'c \& Nakar 2005) so that
$\epsilon_B$ may be even lower. According to Panaitescu \& Kumar
(2002) and Yost et al. (2003), the observed spectra and light
curves of the GRB afterglows imply
 $\epsilon_B\sim 10^{-3}\div 10^{-1}$ in most cases. Eichler \& Waxman
(2005) demonstrated that the above estimates may be rescaled such
that the observations are fitted with values of $\epsilon_B$ that
are smaller by an arbitrary factor $f$, $m_e/m_p<f<1$. Taking this
into account one can see that the Weibel instability could provide
the necessary field unless the field decay is too strong. On the
other hand, as the global structure of the shock transition is
dictated by the Larmor rotation of the protons in the background
field, some new physics could come into play.

A presumably important physical mechanism is the synchrotron maser
instability at the shock front. This instability generates
semi-coherent, low-frequency electromagnetic waves (Gallant et al.
1992; Hoshino et al. 1992; Lyubarsky in preparation). In low
magnetized flows, the amplitude of these waves exceeds the
amplitude of the shock compressed background field. In this case,
relativistic particles radiate in the field of the waves via
nonlinear Compton scattering (e.g., Melrose 1980, pp. 136-141).
The power and characteristic frequencies of this emission are
similar to those for synchrotron emission in the magnetic field of
the same strength therefore the observed gamma ray bursts
afterglows may be attributed to the nonlinear Compton scattering
off the electromagnetic waves generated by the synchrotron maser
instability at the shock front. It is beyond the scope of this
paper to redo afterglow theory with the spectrum of low frequency
electromagnetic waves that is expected from this instability.

We thank J.Frederiksen, A. Spitkovsky and R.Blandford for helpful
discussions. Y.L. acknowledges support from the German-Israeli
Foundation for Scientific Research and Development. 
D.E. acknowledges the hospitality of the Kavli Institute for
Particle Astrophysics and support from the Israel-U.S. Binational
Science Foundation, the Israeli Science Foundation, and the Arnow
Chair of Physics.

\end{document}